\documentclass[prb,twocolumn,aps,amsmath,amssymb,floatfix,showpacs,longbibliography,superscriptaddress]{revtex4-1}

\usepackage{color}
\usepackage{graphicx}
\usepackage{dcolumn}
\usepackage{bm}
\usepackage{subfigure}
\usepackage{array}
\newcolumntype{L}{>{\centering\arraybackslash}m{0.33\textwidth}}
\usepackage[colorlinks=true,linkcolor=blue,citecolor=blue,urlcolor=blue]{hyperref}

\graphicspath{{./figs/}}
\begin{document}

\newcommand{\bo}{\boldsymbol}
\newcommand{\boq}{\mathbf{q}}
\newcommand{\bok}{\mathbf{k}}
\newcommand{\bor}{\mathbf{r}}
\newcommand{\boG}{\mathbf{G}}
\newcommand{\boR}{\mathbf{R}}
\newcommand\2{$_2$}

\newcommand\theosmarvel{Theory and Simulation of Materials (THEOS), and National Centre for Computational Design and Discovery of Novel Materials (MARVEL), \'Ecole Polytechnique F\'ed\'erale de Lausanne, CH-1015 Lausanne, Switzerland}
\newcommand\geneva{Department of Quantum Matter Physics, University of Geneva, CH-1211 Geneva, Switzerland}
\newcommand\nanomat{nanomat/QMAT/CESAM and European Theoretical Spectroscopy Facility, University of Liege (Uliege), Belgium}
\newcommand\modena{Dipartimento di Fisica Informatica e Matematica, Universit\`a di Modena e Reggio Emilia, Via Campi 213/a, I-41125 Modena, Italy}

\title{Profiling novel high-conductivity 2D semiconductors}

\author{Thibault Sohier}
\affiliation{\nanomat}
\affiliation{\theosmarvel}
\author{Marco Gibertini}
\affiliation{\modena}
\affiliation{\geneva}
\affiliation{\theosmarvel}
\author{Nicola Marzari}
\affiliation{\theosmarvel}

\date{\today}


\begin{abstract}
When complex mechanisms are involved, pinpointing high-performance materials within large databases is a major challenge in materials discovery. We focus here on phonon-limited conductivities, and study 2D semiconductors doped by field effects. Using state-of-the-art density-functional perturbation theory and Boltzmann transport equation, we discuss 11 monolayers with outstanding transport properties.
These materials are selected
from a computational database of exfoliable materials providing monolayers that are dynamically stable and that do not have more than 6 atoms per unit cell.
We first analyze electron-phonon scattering in two well-known systems: electron-doped InSe and hole-doped phosphorene. Both are single-valley systems with weak electron-phonon interactions, but they represent two distinct pathways to
fast transport: a steep and deep isotropic valley for the former and strongly anisotropic electron-phonon physics for the latter.
We identify similar features in the database and compute the conductivities of the relevant monolayers. This process yields several high-conductivity materials, some of them only very recently emerging in the literature (GaSe, Bi$_2$SeTe$_2$, Bi$_2$Se$_3$,  Sb$_2$SeTe$_2$), others never discussed in this context (AlLiTe$_2$, BiClTe, ClGaTe, AuI). Comparing these 11 monolayers in detail, we discuss how the strength and angular dependency of the electron-phonon scattering drives key differences in the transport performance of materials despite similar valley structure. We also discuss the high conductivity of hole-doped WSe$_2$, and how this case study shows the limitations of a selection process that would be based on band properties alone.
\end{abstract}

\maketitle

\section{Introduction}

Two-dimensional (2D) semiconductors with excellent intrinsic transport properties would be beneficial to many applications \cite{Butler2013,Akinwande2014,Chhowalla2016a}. Some well-known 2D materials like transition-metal dichalcogenides (TMDs), phosphorene or silicene have been extensively studied both experimentally (MoS$_2$\cite{Radisavljevic2013b,Radisavljevic2011}, Si\cite{Li2019,Li2014}\cite{Tao2015}) and theoretically (MoS$_2$\cite{Li2013,Li,Kaasbjerg2012a,Gunst2016a,Sohier2018}, P$_4$\cite{Rudenko2016,Trushkov2017,Gaddemane2018,Sohier2018}).
Other candidates\cite{Wang2018a} have been proposed using approximate deformation potential models and the Takagi\cite{Takagi1994} formula, but it has been shown that such approaches suffer from limited reliability \cite{Gaddemane2019}.
The full potential of 2D materials for future device applications could be much broader than the dozen materials currently under extensive experimental investigation, since many more monolayers have been predicted to be either exfoliable from experimentally known layered materials\cite{Cheon2017,Ashton2017,Choudhary2017,Mounet2018} or synthesizable\cite{Haastrup2018,Zhou2019}. Such computational collections of prospective 2D materials would likely contain some promising candidates for electronic transport in various contexts.

Finding the ideal material for a given device would require the cross-examination and co-optimization of many different properties \cite{Klinkert2020}.
Here, rather than focusing on one particular application, we are interested in the physical features leading to good transport performance. We describe the tools and knowledge needed to explore large databases with an informed, purposeful approach, and eventually identify the candidates with the highest conductivity.

We focus here on room-temperature phonon-limited electronic transport and search across 2D semiconductors. First-principles methods~\cite{Giustino2017} have proven quite useful and successful in predicting physics and properties in this regime\cite{Ponce2020} and for 2D materials \cite{Park2014,Sohier2014a}, provided one is aware of its limits\cite{Gaddemane2019}. Although dedicated codes exist\cite{Ponce2016,Zhou2020}, performing state-of-the-art first-principles transport simulations on large sets of materials remains nevertheless a challenge, and few works tackle more than one material at a time. Yet, an overall panorama on the property landscape would be very valuable to materials design. For example, in a previous work involving 5 materials \cite{Sohier2018}, we have shown the importance of intervalley scattering, while the authors of Ref.~\onlinecite{Cheng2019} have studied 4 hexagonal elemental materials to show the benefits of a sharp and deep single valley.

In this work, we explore our portfolio of exfoliable 2D materials\cite{Mounet2018} taken from the Materials Cloud \cite{MC2D,Talirz2020}, limiting ourselves to at most 6 atoms per unit cell. We select the most promising systems from a band structure analysis, and then study their transport properties using an accurate and automated framework that we recently developed\cite{Sohier2018}, finding several excellent 2D semiconductors. Compared to other methods\cite{Brunin2020,Ponce2020}, the approach of Ref.~\onlinecite{Sohier2018} has two main advantages in the context of this work: i) it includes several tools to analyze band structure properties like valley structure or Fermi velocity and ii) the calculation of conductivity is automated within the AiiDA framework\cite{Pizzi2016,Huber2020}, which is key to study many materials in a high-throughput fashion. These tools are available\cite{MCArchive} to the community on the Archive section of the Materials Cloud \cite{Talirz2020}, in support of the FAIR principles of open science and open data.

High carrier densities are routinely obtained in monolayer 2D semiconductors by field-effect doping, especially using ionic-liquid gating\cite{Zhang2012,Braga2012,Ye2012}. Here, we consider a fixed carrier density of $n/p=10^{13}$ cm$^{-2}$ (electrons or holes). Electron-phonon interactions are computed in an electrostatic framework including such doping\cite{Sohier2017}, as well as screening from the induced free electrons or holes. This is in contrast to other first-principles studies, usually simulating electron-phonon interactions (EPI) in neutral materials, which would be valid only in the limit of vanishing carrier density. The capability to study explicitly the high-doping regime is a valuable complement to this, and very relevant for monolayers operating in the degenerate limit, when the chemical potential is close to or inside the conduction or valence band (doping regimes are further discussed in App.~\ref{sec:doping}). In addition, its predictive accuracy provides meaningful comparisons to experiments. Indeed, the high-doping regime is often used experimentally to characterize the intrinsic properties of novel materials, because it allows to screen charged impurities.

In this study we find excellent phonon-limited transport properties for electron-doped Bi$_2$SeTe$_2$, Bi$_2$Se$_3$, BiClTe, Sb$_2$SeTe$_2$, InSe, GaSe, AlLiTe$_2$ and hole-doped phosphorene (P$_4$), AuI, ClGaTe, and WSe$_2$, all showing mobilities in the range from few hundreds to few thousands cm$^2$/Vs. Three of these 2D materials are very well-known and studied (P$_4$\cite{Rudenko2016,Trushkov2017,Gaddemane2018,Sohier2018}, WSe$_2$\cite{Sohier2018}, InSe\cite{Li2019a}). The Sb$_2$X$_3$ and Bi$_2$X$_3$ (X = Se, Te, S) compounds are better known for the topological properties of their 3D parents\cite{Zhang2010,Jafarpisheh2020}.
Monolayers have been recently studied in the context of phonon-limited transport \cite{Liu2020,Wang2018a,Tang2020}, but only within the approximate Tagaki formalism~\cite{Takagi1994}.
The transport performance of monolayer Sb$_2$SeTe$_2$ has been confirmed experimentally\cite{Qu2019}. GaSe (along with InSe) was studied\cite{Chen2019b} ab initio with a representation of electron-phonon scattering that goes beyond deformation potential theory while this work was carried out. AlLiTe$_2$, BiClTe, ClGaTe, and AuI are, to our knowledge, still new in the context of 2D charge transport. Beyond the relatively large number of materials identified and the novelty of some of them, this work most importantly provides visual and intuitive understanding of electron-phonon scattering as well as a systematic and data-supported analysis of the features leading to good transport performance. We confirm and extend previous remarks on the importance of band properties such as number of valleys, effective masses or anisotropy; and show that these can be used for larger scale databases studies to shortlist prospective conductors. On the other hand, we also show how differences in the strength and angular dependency of electron-phonon scattering still induce significant variations in the conductivity of materials with similar band properties.

\section{Computational and theoretical framework}

\subsection{Computational details}

First-principles calculations of structures, bands, phonons and electron-phonon interactions are performed with the Quantum ESPRESSO\cite{Giannozzi2009,Giannozzi2017} (QE) software, by combining density-functional theory (DFT) and density-functional perturbation theory (DFPT) within the generalized gradient approximation as formulated by Perdew, Burke, and Ernzerhof \cite{PBE} (PBE). 2D periodic-boundary conditions are applied and the electrostatics of a symmetric double-gate field-effect setup are simulated using the approach described in Ref.~\onlinecite{Sohier2017}. Open-boundary conditions are  important to properly describe polar-optical phonons\cite{Sohier2017a} and screening\cite{Sohier2015}. The field-effect setup is used to induce an electron or hole density with a default value of $n/p = 10^{13}$ cm$^{-2}$. The simulation of this relatively high-doping regime is further discussed in App.~\ref{sec:doping}. Calculations are managed using the AiiDA materials informatics infrastructure\cite{Pizzi2016,Huber2020}. The AiiDA database containing the provenance for the transport calculations and the tools necessary to reproduce this work are provided in the Archive section of the Materials Cloud \cite{MCArchive,Talirz2020}.
The standard numerical setup for all ground state and phonon calculations has $32\times32$ Monkhorst-Pack k-point grids to sample the full Brillouin zone, 0.02 Ry cold smearing\cite{Marzari1999} with SSSP pseudopotentials \cite{Prandini2018a} (efficiency version 0.7) and energy cutoffs recommended therein. PAW pseudopotentials were substituted because they are incompatible with the use of symmetry in the electron-phonon routines of QE. Other minor variations from the standard setup can be found in the AiiDA database  provided.
The use of cold smearing allows k-point convergence while keeping a lower effective temperature in the calculations, making the free-carrier screening closer to what it would be in real conditions.
Spin-orbit interactions are included only for hole-doped WSe$_2$, where they play a significant role. For other materials, while spin-orbit interactions may have an effect on the band structure in general (e.g. on the band gap for BiSeTe$_2$ \cite{Wang2018a}), they will not have large consequences on the conductivity or mobility as long as there is no significant energy splitting of valleys with opposite spins (small variations may come from changes in the effective masses).
The analysis starts from a database of band structures computed (non-self-consistently) in the neutral material on very fine electronic momenta grids (about 90 by 90) to analyse the valley structure. These bands are then recomputed with field-effect doping for the selected materials. Phonon momenta are chosen to include only relevant transitions, and the Boltzmann transport equation (BTE) is solved as described in Ref.~\onlinecite{Sohier2018}. We improved the stability for the solution of the BTE by using the velocity of each initial state as a fictitious electric field direction [Eq. (11) of Ref.~\onlinecite{Sohier2018}].

\subsection{Boltzmann transport equation}

The solution of the  BTE is briefly outlined here for convenience; more details can be found in Ref. \onlinecite{Sohier2018}. The (longitudinal) conductivity is computed as follows:
\begin{align} \label{eq:conductivity}
\sigma=\frac{1}{\rho}=  2e^2 \int  \frac{d\bok}{(2\pi)^2}
\left( \bo{v}(\bok) \cdot \bo{u}_{\bo{E}}\right)^2 \tau(\bok)\left[-\frac{\partial f^0(\varepsilon_{\bok})}{\partial \varepsilon}\right]
\end{align}
where $e$ is the Coulomb charge, and $\bok,\varepsilon_{\bok},\bo{v}(\bok)$ represent electronic momenta (and band index implicitly), energies, and velocities, respectively. The unit vector $\bo{u}_{\bo{E}}$ points in the electric field's direction and $f^0$ is the Fermi-Dirac distribution.
Here $\tau$ is an energy- and momentum-dependent variable that has the dimensions of time and solves the linearized BTE:
\begin{align} \label{eq:BTE}
\begin{split}
(1-f^0(\bok))  \bo{v}(\bok) \cdot \bo{u}_{\bo{E}} = & \sum_{\bok'} P_{\bok\bok'} (1-f^0(\bok')) \times  \\
& \left\{ \bo{v}(\bok) \cdot \bo{u}_{\bo{E}} \tau(\bok)  -  \bo{v}(\bok') \cdot \bo{u}_{\bo{E}} \tau(\bok')
\right\}
\end{split}
\end{align}
$P_{\bok\bok'}$ being the probability for an electron in state $\bok$ to be scattered into state $\bok'$. In the following we consider only  phonon-induced scattering, so that:
\begin{align}\label{eq:Pkkp}
\begin{split}
P_{\bok \bok+\boq}= \sum_{\nu} \frac{2\pi}{\hbar} \frac{1}{N} & |g_{\bok \bok+\boq, \nu}|^2
\{
n_{\boq, \nu} \delta(\varepsilon_{\bok+\boq}-\varepsilon_{\bok}-\hbar \omega_{\boq, \nu}) \\
+&(n_{\boq, \nu}+1) \delta(\varepsilon_{\bok+\boq}-\varepsilon_{\bok}+\hbar \omega_{\boq, \nu})
\}.
\end{split}
\end{align}
where $\boq,\nu,\hbar \omega_{\boq, \nu}, n_{\boq, \nu}$ are phonon momenta, mode index, energy and occupations, and $g_{\bok \bok+\boq, \nu}$ are the electron-phonon coupling matrix elements.

\section{High-conductivity 2D semiconductors}

\begin{figure}[h]
  \includegraphics[width=0.45\textwidth]{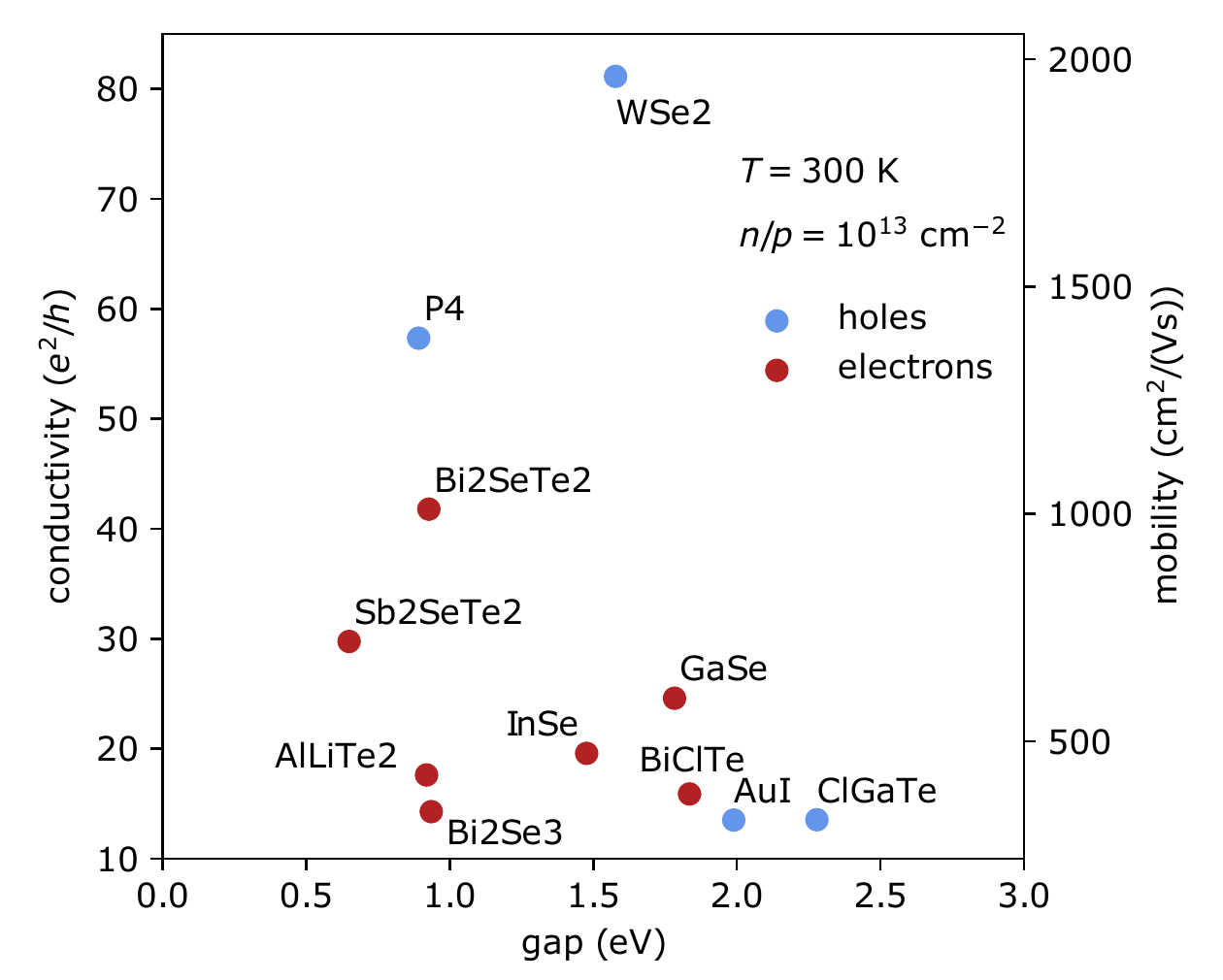}
  \caption{Computed conductivities and mobilities at room temperature plotted against PBE gap for all the materials considered in this work. BTE is solved for a fixed carrier density of $n/p=10^{13}$ cm$^{-2}$ (electrons or holes).}
  \label{fig:conductivities}
\end{figure}

We study in careful detail eleven monolayers obtained from our database of 256 easily exfoliable materials with at most 6 atoms per unit cell \cite{Mounet2018}, as available on the Materials Cloud\cite{MC2D,Talirz2020} and in the supplementary material of Ref.~\onlinecite{Mounet2018}. The corresponding conductivity and mobility, as a function of the PBE gap, are shown in Fig.~\ref{fig:conductivities}. At this doping level a conductivity of a few $e^2/h$ is already considered good \cite{Radisavljevic2013b,Tao2015} for a semiconductor, while the best graphene devices yield values around $500$ $e^2/h$ \cite{Wang2013a,Banszerus2019}. The materials presented in this work are in the intermediary orders of magnitude, with conductivities of $10 \sim 100$ $e^2/h$, and mobilities in the same range as bulk silicon ($400$ cm$^2$/Vs for holes, $1400$ cm$^2$/Vs for electrons).

In the following, we detail the exploration process that led to those materials. First, two well-known high-conductivity 2D semiconductors are analyzed to identify representative band features. We then search for those features within the Materials Cloud database\cite{Mounet2018,MC2D} and compute the transport properties of the most promising candidates. Electron-phonon scattering is analyzed in each monolayer. The suffix ``-e'' or ``-h'' is attached to the materials' formula to indicate if either electron doping or hole doping is considered.

\subsection{Prototypical high-conductivity materials: InSe and P$_4$}

We start by analyzing key features of two well-known semiconductors with excellent transport performance, InSe on the electron side (InSe-e) and phosphorene on the hole side (P$_4$-h). In the following we will use repeatedly two different plots to visualize, understand and compare the details of the transport properties of 2D materials, reported first in Fig.~\ref{fig:scatt_plots_ref} for InSe and P$_4$.
On the left, a ``velocity plot'' shows the band structure in a format that is relevant for transport. For each electronic state $\bok$ (each black dot) on the fine k-point grid used for solving the BTE and within a certain energy range from the band edge (set as the origin of the y-axis), the energy $\varepsilon_{\bok}$ is plotted against the norm of the velocity $|\bo{v}|=|\frac{1}{\hbar} \nabla_{\bok} \varepsilon_{\bok}|$, given in atomic Rydberg units (ARU)~\footnote{Atomic Rydberg units are the most convenient units for velocity in materials. Typical Fermi velocities in the database --including graphene-- are on the order of 1 ARU, which corresponds to $\sim 10^{6}$~m/s.}.
The spread indicates the anisotropy of the valley. The red scale of the background is proportional to the derivative of the Fermi-Dirac occupation at $300$~K, thus highlighting the states participating to transport with a non-vanishing contribution to Eq.~\eqref{eq:conductivity}. The Fermi level is computed at room temperature (i.e. such that $\int_{\bok} f^0(\varepsilon_{\bok}, T) d\bok/(2\pi^2) = 10^{13}$ cm$^{-2}$ for electrons, and similarly for holes).
On the right side, a ``scattering plot'' shows where and how easily a certain initial state can be scattered. The initial state at $\bok_{in}$, indicated by a black square, is chosen to be at the Fermi level (and in the transport direction, when relevant). The grey shading shows the morphology of the valley(s) considered for transport.
The red color scale represents an effective coupling constant $g_{eff}$ which accounts for all phonon modes, their occupation at $300$ K, and energy conservation, since this quantity can be meaningfully compared between materials. In practice, $g_{eff}$ is defined as the square root of the sum  of the interpolated
$|g_{\bok_{in}, \bok_{in}+\bo{q}, \nu }|^2 n_{\boq, \nu}$ ( or $n_{\boq, \nu}+1$) over all $\nu$ and $\bo{q}$ that fulfill energy conservation during phonon absorption (or emission) when the final state at $\bok_{in}+\bo{q}$ falls into a certain zone (the valley being tessellated into triangles). The value given at the top of each scattering plot is the lifetime of the initial state computed within the momentum relaxation time approximation:
\begin{equation}\label{eq:mRTAtau}
\frac{1}{\tau(\mathbf{k})} = \sum_{\mathbf{k}'} P_{\mathbf{k}\mathbf{k}'} \frac{1-f^0(\mathbf{k}')}{1-f^0(\mathbf{k})} \times \left\{1-  \frac{\mathbf{v}(\mathbf{k}') \cdot \mathbf{v}(\mathbf{k}) }{\mathbf{v}(\mathbf{k})^2}
\right\}
\end{equation}
Everything is computed at 300~K, with the corresponding chemical potential in $f^0$.
\\

We now discuss qualitatively the features allowing electron-doped InSe and hole-doped phosphorene to maximize the conductivity in Eq.~\eqref{eq:conductivity}:
\begin{figure*}
  \includegraphics[width=0.9\textwidth]{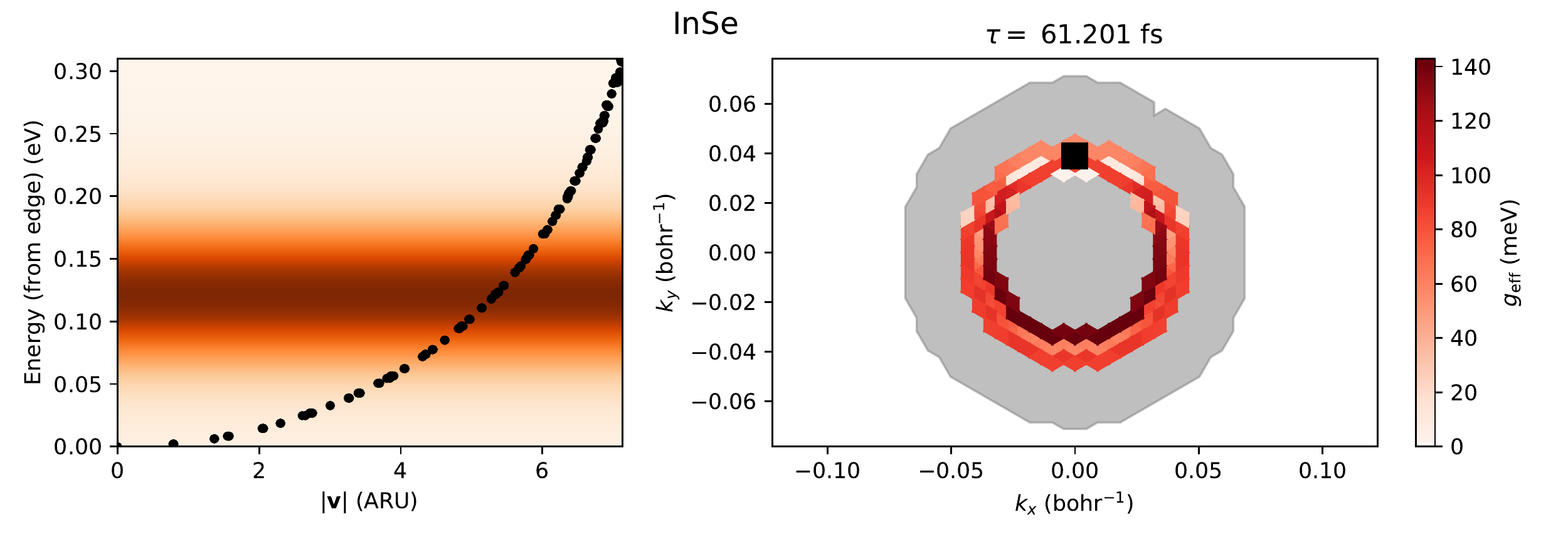}
  \includegraphics[width=0.9\textwidth]{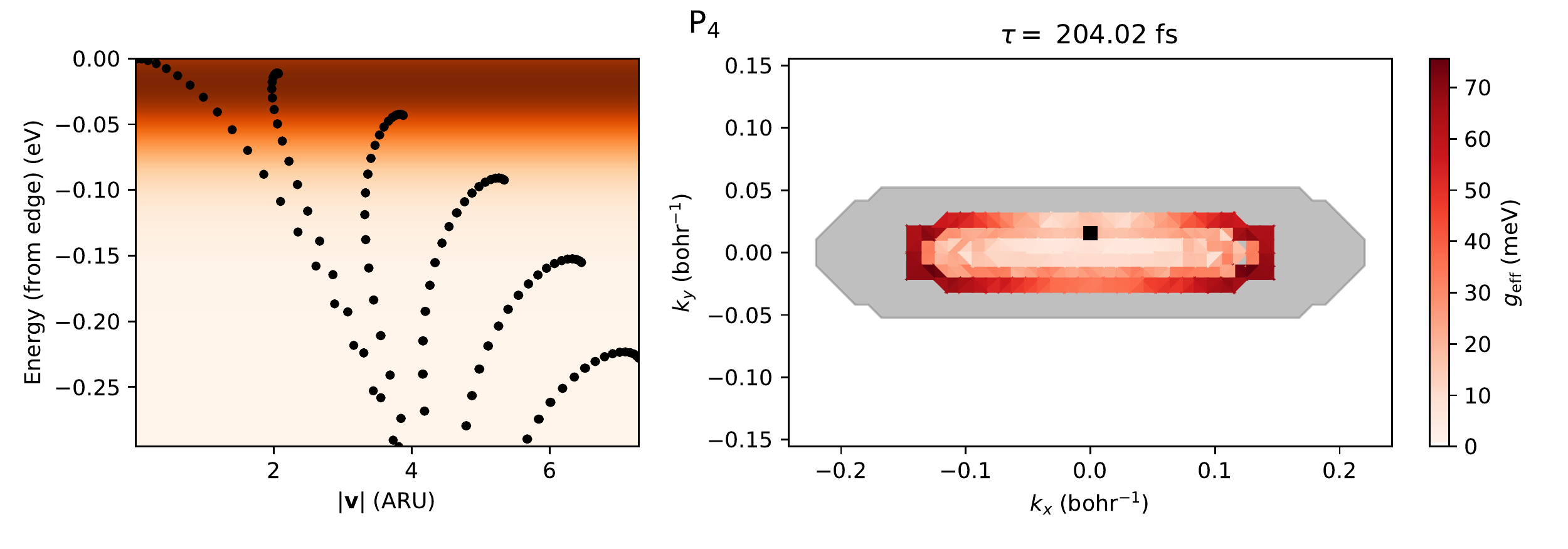}
  \caption{Transport properties of InSe and phosphorene (P$_4$), two well-known good conductors. On the left, the ``velocity plot'' shows, for each electronic state $\bok$ in the valley, the energy $\varepsilon_{\bok}$ from the band edge plotted against the norm of the velocity $|\bo{v}|=|\frac{1}{\hbar} \nabla_{\bok} \varepsilon_{\bok}|$, in atomic Rydberg units (ARU).
  The color scale of the background is proportional to the derivative of the Fermi-Dirac distribution, which appears in Eq.~\eqref{eq:conductivity}. On the right, the ``scattering plot'' shows where an initial state (black square) can be scattered within the valley (grey shade). The color scale represents the effective coupling constant $g_{eff}$, which accounts for all phonon modes, their occupation at $300$ K, and energy conservation. At the top, $\tau$ is the scattering time of the initial state computed within the momentum relaxation time approximation, Eq.~\eqref{eq:mRTAtau}. }
  \label{fig:scatt_plots_ref}
\end{figure*}
this is a sum over electronic states weighted by the derivative of the Fermi-Dirac occupation. The weight effectively selects states around the chemical potential within an energy range that scales with temperature, as represented in the left panels of Fig.~\ref{fig:scatt_plots_ref}.
The rest of the integrand can be separated in two contributions: i) one from scattering (electron-phonon here) through the scattering lifetime $\tau$; ii) and one from the velocity of the carriers $\bo{v}$ projected along the transport direction $\bo{u_E}$.

The scattering contribution i) is inversely proportional to the strength of the EPI and the amount of states available for scattering (as apparent in Eq.~\eqref{eq:mRTAtau}), i.e. the density of states (DOS) within a certain energy window around the initial states, depending on the energy of the phonons involved. We note, however, that the DOS contribution is essentially compensated when doing the integral over electronic states to get the conductivity, Eq.~\eqref{eq:conductivity}. To maximize $\tau$, the quantity to focus on is thus the strength of the EPI, which should of course be minimized. Weak EPIs is a feature shared by InSe and phosphorene.

There are different ways to maximize the velocity contribution ii); in general, one needs to maximize the velocity in the direction of transport.
In the case of InSe, the velocity is isotropic and very large, with a small effective mass. Accordingly, the DOS is low and the Fermi level reaches high energies above the valley edge, towards higher velocities, eventually saturating at a maximal value in case of non-parabolic materials. The benefits of a such a steep and deep single valley is that current carrying states around the chemical potential have high velocities, as also pointed out in Ref.~\onlinecite{Cheng2019}.
In the contrasting case of phosphorene, the valley is highly anisotropic. Provided one chooses the low effective mass direction for transport, the projection of the velocity in the transport direction is maximized. However, the DOS here is relatively larger, and the Fermi level stays closer to the band edge, where  velocities are lower. The good performance of phosphorene thus relies on a somewhat more fragile balance.

An essential feature to maximize both i) and ii) is to work with single valley materials.
Single valley band structures lead to a lower DOS, higher Fermi level and higher velocities. Even more compelling, one avoids intervalley scattering, which is usually stronger than intravalley scattering and hinders transport \cite{Sohier2018}.

In a related work\cite{Klinkert2020}, several promising materials for ultra-short transistor devices have been identified.  In that situation, one optimizes the performance of the material in the ballistic limit and the role of electron-phonon scattering is secondary, since the channels are usually below the scattering length. Thus, maximizing both the velocity in the transport direction and the DOS is extremely beneficial, and anisotropic and multivalley materials are interesting. When electron-phonon scattering is turned on, multivalley band structures are not ideal, while anisotropy can remain relevant.

\subsection{More like InSe: Steep and deep single-valleys}

\begin{figure*}
 \includegraphics[width=0.9\textwidth]{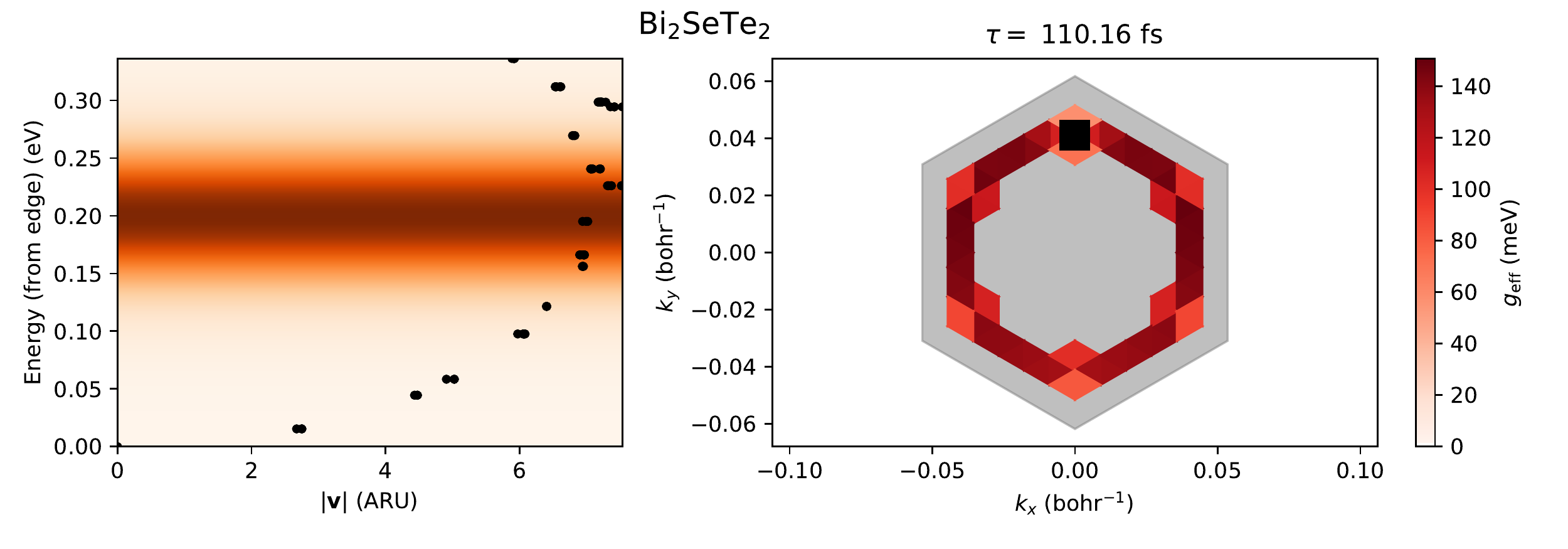}
 \includegraphics[width=0.9\textwidth]{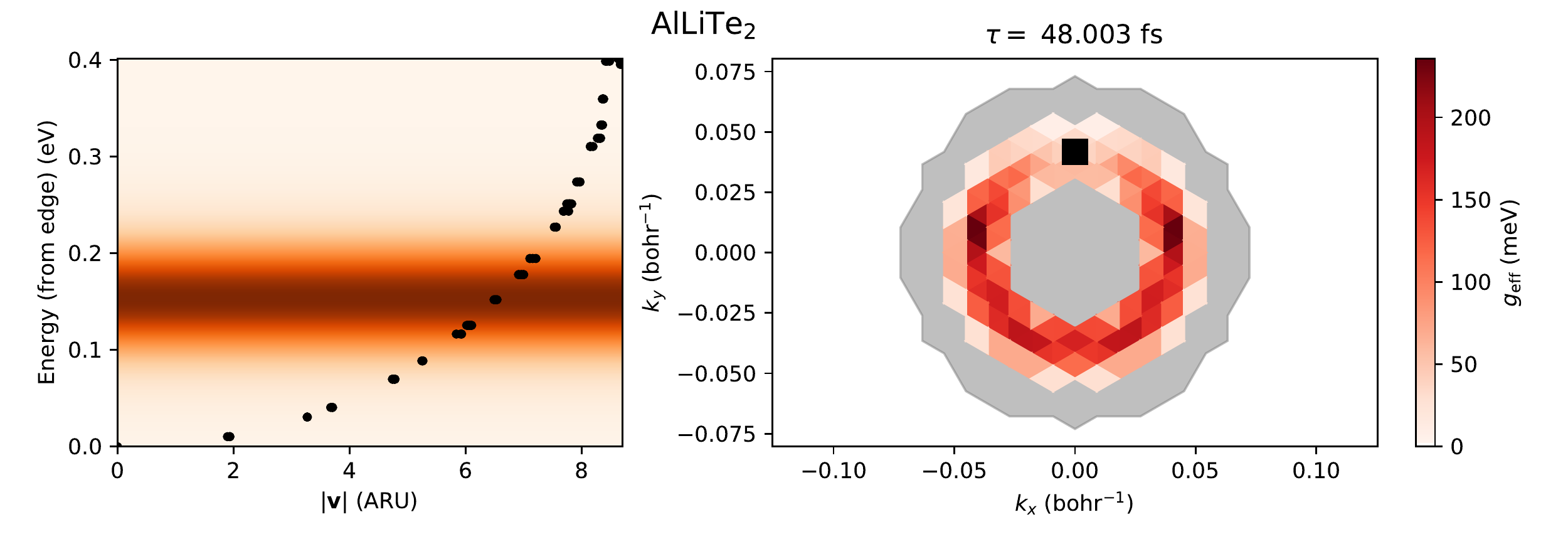}
  \caption{Transport properties of Bi$_2$SeTe$_2$ ($\sigma = 42~e^2/h$) and AlLiTe$_2$ ($\sigma = 18~e^2/h$), showing the velocity (left) and scattering (plots) as described in Fig.~\ref{fig:scatt_plots_ref}.  Both monolayers belong to a category of materials with steep and deep single valleys in the conduction band. Similar plots are given in App.~\ref{app:add_data} for other monolayers in this category: BiClTe-e, GaSe-e, Bi$_2$STe$_2$-e, Bi$_2$Se$_3$-e, and Sb$_2$SeTe$_2$-e.}
  \label{fig:scatt_plotse}
\end{figure*}

As shown in Fig.~\ref{fig:scatt_plots_ref}, InSe  has a sharp and deep single electron valley, with carrier velocities close to  $6$ ARU at the Fermi level when $n=10^{13}$~cm$^{-2}$ (roughly 6 times larger than the Fermi velocity in graphene, $\sim 1$~ARU).
Steep isotropic valleys are obviously characterized by low DOS, which means that the Fermi level quickly shifts away from the band edge with increasing carrier density. This allows to reach higher carrier velocities, but it often means that the Fermi level reaches other valleys in the band structure.
In that case, the aforementioned benefits of the single valley structure are lost. If the Fermi level is close to the edges of the next valleys, intervalley scattering is activated and the new states populated are not helping to conduct due to their low velocity. Note that, in a similar spirit, the suppression of intervalley scattering via valley-engineering (e.g.\ through strain) has been proposed\cite{Sohier2019} to enhance the transport properties of multi-valley materials such as arsenene.
Thus, a steep valley is not sufficient to have good transport performance: it must also be deep. More precisely, higher-energy valleys need to be far enough for the material to operate effectively as a single valley material for a doping of $10^{13}$ cm$^{-2}$.

We look for such valleys among the exfoliable materials in our study set. We use band structures computed on very fine grids in the neutral materials and select materials with positive phonons, a limited gap $E_g<2.5$ eV, a single valley (within 100 meV of the Fermi level), and a maximum Fermi velocity $|\bo{v}_{\rm{max}}|> 6$ ARU.  This allows us to identify BiClTe-e, AlLiTe$_2$-e, GaSe-e, Bi$_2$SeTe$_2$-e, Bi$_2$STe$_2$-e, Bi$_2$Se$_3$-e, and Sb$_2$SeTe$_2$-e as promising candidates.
Electron-phonon scattering is studied for these monolayers (except Bi$_2$STe$_2$-e,  a bit redundant with respect to the other two members of the Bi$_2$X$_3$ family). The results are shown in Fig.~\ref{fig:scatt_plotse} for Bi$_2$SeTe$_2$-e and AlLiTe$_2$-e, and in the appendix for the rest. The general similarity of the plots indicates at a glance why these materials have similar transport properties. The carrier velocities at the Fermi level, the strength of EPI, and the scattering times are of similar orders of magnitude. A closer look reveals the reasons underlying their precise ranking.
BiSeTe$_2$-e, BiSe$_3$-e, and Sb$_2$SeTe$_2$-e are very similar in terms of chemistry, band structure, and phonons. BiSeTe$_2$-e (see Fig.~\ref{fig:scatt_plotse}) displays the highest conductivity ($\sigma = 42~e^2/h$), combining some of the weakest EPI with the largest velocities. It is followed by Sb$_2$SeTe$_2$-e ($\sigma = 30~e^2/h$) with similar EPI but slightly smaller velocities. Bi$_2$Se$_3$-e  ($\sigma = 14~e^2/h$) also has a very sharp valley, but the EPI is $2$ to $3$ times stronger.
As could be expected, electron-phonon scattering in GaSe-e ($\sigma = 25~e^2/h$) is similar to InSe-e ($\sigma = 20~e^2/h$), with scattering times differing by only $\sim10\%$. GaSe-e owes its higher conductivity mostly to a sharper valley and higher velocities.
Finally, AlLiTe$_2$-e ($\sigma = 18~e^2/h$, see Fig. \ref{fig:scatt_plotse}) and BiClTe-e ($\sigma = 16~e^2/h$) have slightly larger velocities but stronger EPIs than InSe-e, making them slightly less conductive.
Note that all these systems with sharp valleys are electron-doped, which may be rationalized by considering that conduction bands are more often made of delocalized non-bonding states which tend to be more dispersive. Despite the valleys being very similar (isotropic with $|\bo{v}_{\rm{max}}| \approx 6 \sim 7$ ARU), the conductivity varies from $14$ to $42$
$e^2/h$. This points to the importance of accounting for the details of EPIs to rank materials accurately.

\begin{figure*}
\includegraphics[width=0.9\textwidth]{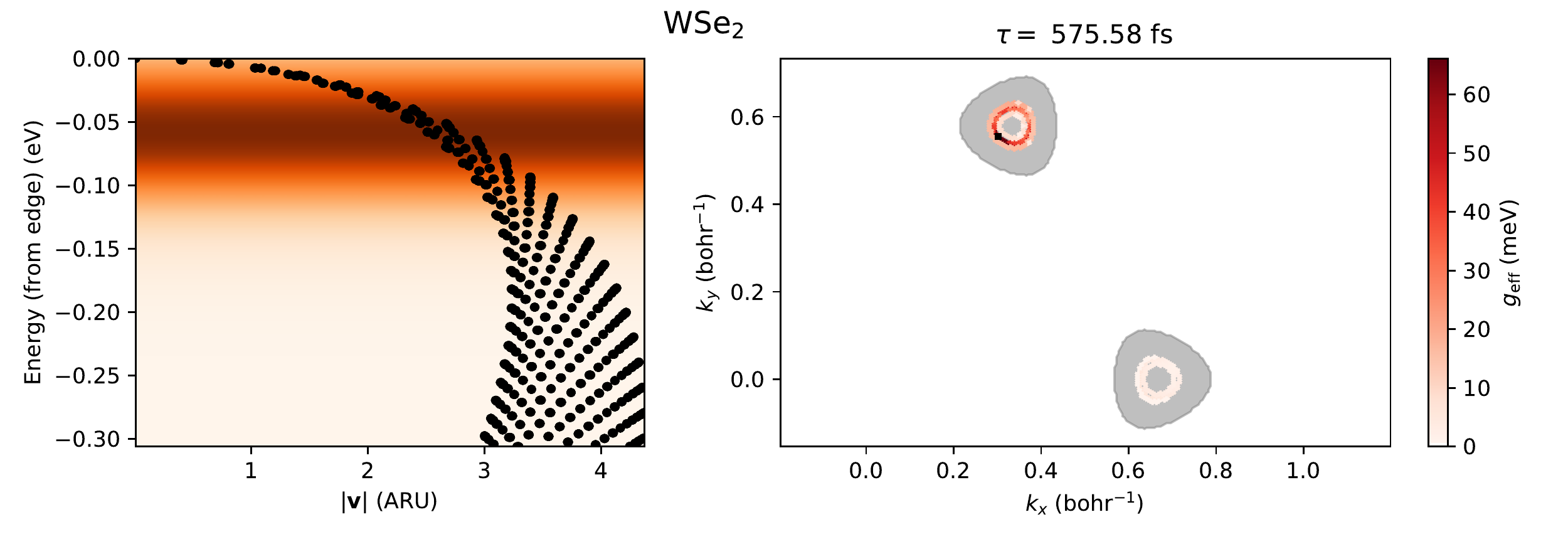}
  \caption{Transport properties of WSe$_2$-h ($\sigma = 81~e^2/h$), showing the velocity (left) and scattering (plots) as described in Fig.~\ref{fig:scatt_plots_ref}. Thanks to spin-orbit interactions, weak spin-flip scattering, and the effective absence of any valley at $\Gamma$, spins travel in parallel channels not connected to each other. }
 \label{fig:scatt_plotsWSe$_2$h}
\end{figure*}

We also note that the selection process does not guarantee to find all the best conductors, as demonstrated by the example of WSe$_2$-h: TMDs were not selected because they are not single-valley (also their Fermi velocities would be too small). However, thanks to strong spin-orbit interactions, the hole side of TMDs can be considered to be in the steep and deep single valley category. Indeed, as is well known, the hole valleys associated to opposite spin textures split very strongly in energy, of the order of 100 meV, making the lower valleys irrelevant for transport. One effectively obtains two valleys at K and K' with opposite spin textures. WSe$_2$-h, at least within our computational framework, also has the advantage that the edge of the valence band at $\Gamma$ is quite low, eliminating a potential intervalley scattering channel. Furthermore, as shown in Fig.~\ref{fig:scatt_plotsWSe$_2$h}, the intervalley  scattering, associated with spin-flip EPIs, is weak ($<10$ meV) compared to intravalley, spin-conserving EPIs ($\sim 50$ meV). This implies that the two valleys are effectively decoupled as far as phonon-limited transport is concerned and transport is similar to the single valley case, except opposite spins travel in separate channels located at different points in the Brillouin zone. Comparing with the isotropic single-valley materials discussed above,  WSe$_2$-h has velocities half as small but EPIs at least 3 times weaker, leading to the largest conductivity ($\sigma = 81~e^2/h$). This high value is due to the (effective) absence of the $\Gamma$ valley. Indeed,  opposite spins are degenerate at $\Gamma$, and if the $\Gamma$ valley were accessible for scattering, intervalley scattering from both K and K' would be quite strong. In addition, as shown in Ref.~\onlinecite{Sohier2019}, multi-valley occupations enhances the intravalley coupling to the homopolar optical phonon mode by making free-carrier screening inefficient. At any rate, the extraordinarily weak EPI in WSe$_2$-h is tied to subtle spin-orbit and screening effects that are difficult to predict and highlights the limitations of simple selection processes based solely on the band structure.

\subsection{More like phosphorene: anisotropic single valleys}

In contrast with the fairly isotropic band structures of InSe, the case of phosphorene showed that anisotropy can also lead to good performance. So,
\begin{figure*}
 \includegraphics[width=0.9\textwidth]{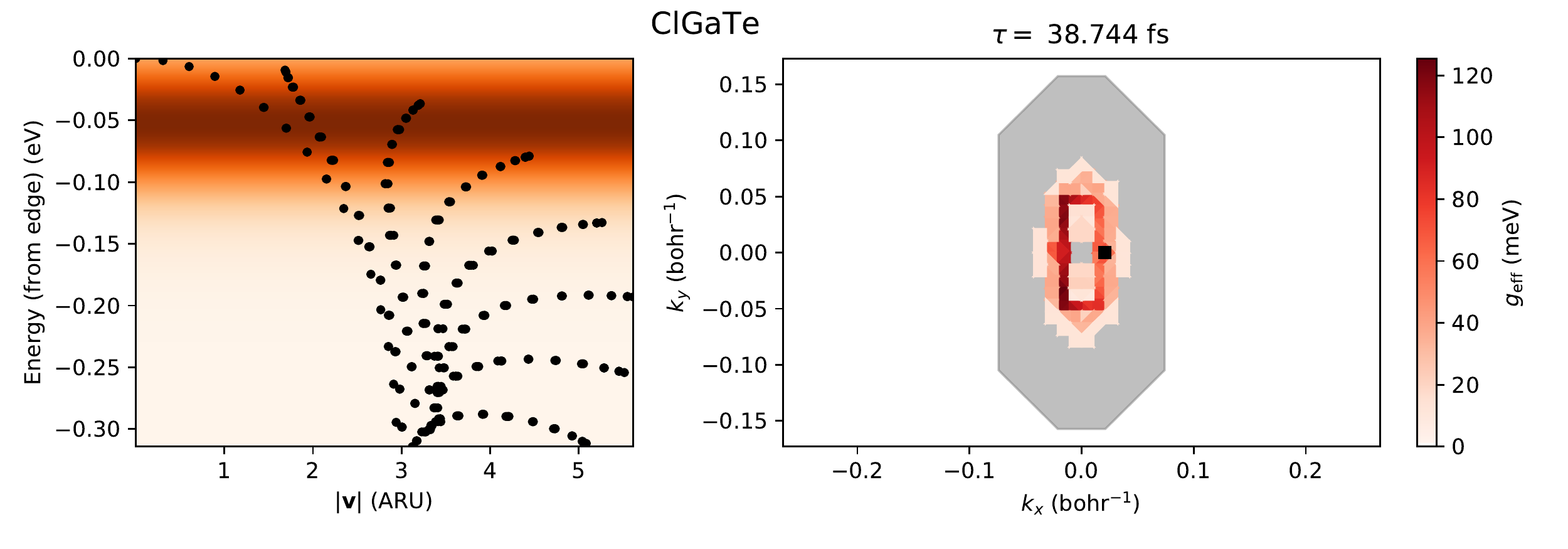}
  \includegraphics[width=0.9\textwidth]{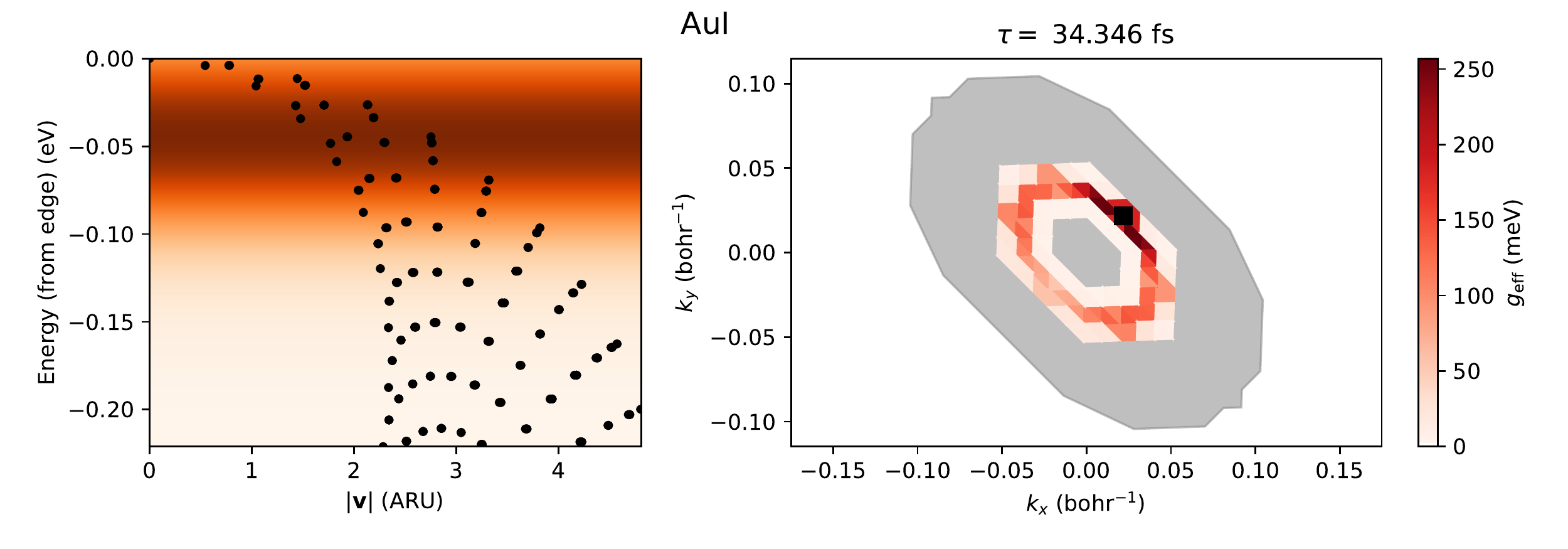}
  \caption{Transport properties of ClGaTe and AuI showing the velocity (left) and scattering (right) plots as described in Fig.~\ref{fig:scatt_plots_ref}. Anisotropic valleys imply there is one optimal transport direction, corresponding to higher velocities. Initial states (marked with a black square) are chosen to be in that direction.}
  \label{fig:scatt_plotsh}
\end{figure*}
we look for similar materials in our database, having stable phonons, single valleys and small band gap at the PBE level ($E_g<2.5$ eV). This time we filter materials keeping materials that combine high velocity ratios ($\frac{v_{max}}{v_{min}}<1.7$) and a decent maximum velocity ($v_{max}>2.0$ ARU) at the Fermi level. These search criteria allow us to identify AuI-h, and ClGaTe-h as well as the electron-side of P$_4$-e as promising candidates.
The electron side of phosphorene is a similar, less pronounced version of the hole side, and was already studied in our previous work\cite{Sohier2018}. Thus, we focus here on ClGaTe-h ($\sigma = 13~e^2/h$) and AuI-h ($\sigma = 14~e^2/h$).

The results of the electron-phonon calculations are plotted in Fig.~\ref{fig:scatt_plotsh}.
The spread of the velocity traces indicates high anisotropy, the velocity varying by a factor 2 at a given energy. The high-velocity direction preferable for transport is $\mathbf{x}$ for ClGaTe and $\mathbf{x}+\mathbf{y}$ for AuI. The anisotropy allows one to benefit from many high velocity states, thanks to a flatter band in the direction perpendicular to transport. However, if the band is too flat and the DOS too high, the Fermi level stays close to the band edges and the velocities are too low.
Here, velocities of the order of $3\sim 4$ ARU ensure good transport properties.
While ClGaTe-h and AuI-h share the anisotropic character of phosphorene, their conductivity remains lower, which in part reflects the fact that the good transport properties of anisotropic materials rely on a more fragile balance of features.

A closer look at the scattering plots in Fig.~\ref{fig:scatt_plotsh}, contrasted with phosphorene in Fig.~\ref{fig:scatt_plots_ref} further reveals the reasons behind phosphorene's superior conductivity. In addition to generally weaker EPIs (which is partly due to the monoatomic nature of P$_4$, eliminating all Born effective charge and piezoelectric couplings), one can observe in P$_4$ a predominance of ``side-scattering'': states in the direction of transport are mostly scattered to states on the sides of the valley with velocities perpendicular to the direction of transport. As can be seen from Eq.~\eqref{eq:mRTAtau}, this leads to a longer scattering times, since the term involving a scalar product of the velocities vanishes.
The contribution of the DOS from the integral of the conductivity usually cancels out with the integral over available scattering states. This is not the case here because we benefit from the larger weight of high velocity states in the conductivity integral while having less dense, low-velocity states in the scattering integral.
Thus,  ``side scattering'' definitely brings the performance of P$_4$-h from great to exceptional.
ClGaTe-h is similar to P$_4$ in terms of velocities, but the EPI is stronger and, importantly, mostly backscattering. The scattering time is consequently 10 times larger. AuI-h has lower velocities and higher EPI.

While targeting the band features of phosphorene has allowed us to find other excellent candidates, we see that the essence of its exceptional performance comes from something much harder to identify at the level of a database: the anisotropy of its electron-phonon interactions. Thus, phosphorene is at the same time both a prototype and an example of the limitations of selection processes based on band structures.

\section{Conclusions}

State-of-the-art density-functional perturbation theory and the Boltzmann transport equation are used to study the outstanding transport properties of several 2D semiconductors. Focusing on conductivity at a fixed density of $n/p = 10^{13}$ cm$^{-2}$, the present results offer a complementary perspective with respect to most first-principles calculations valid in the zero carrier density limit.
We provide a detailed analysis of electron-phonon scattering in two well-known high-conductivity systems: electron-doped InSe and hole-doped phosphorene. While they share some features --like weak EPI and a single-valley electronic structure-- they exemplify two different strategies to maximize the conductivity. InSe's high-velocity, isotropic valley can be exploited thanks to the fact that the next valleys are much higher in energy. Phosphorene, instead, owes its excellent transport performance to the anisotropy of both its band structure and electron-phonon scattering.
Analyzing the band properties of around $\sim 150$ small stable semiconductors with 6 atoms or less in the  unit-cell, from the Materials Cloud, we identify systems with band features similar to either InSe or phosphorene.  We find large phonon-limited conductivities for electron-doped Bi$_2$SeTe$_2$, Bi$_2$Se$_3$, BiClTe, Sb$_2$SeTe$_2$, AlLiTe, and GaSe, as well as hole-doped AuI, ClGaTe, and WSe$_2$. These results confirm that the band structure landscape plays an important role in determining transport and shows that seeking peculiar features in the electronic structure does lead to high-performance materials. Nevertheless, we also show how the details of the strength and angular dependency of electron-phonon scattering play a critical  role in ranking those materials with respect to each other.

\section*{Acknowledgements:}
The authors are grateful to Davide Campi for sharing the initial band structures. This work has been in part supported by NCCR MARVEL. Simulation time was awarded by PRACE on  Marconi at Cineca, Italy (project id.\ 2016163963). Computational resources have been provided by the Consortium des Équipements de Calcul Intensif (CÉCI), funded by the Fonds de la Recherche Scientifique de Belgique (F.R.S.-FNRS) under Grant No. 2.5020.11 and by the Walloon Region. T.S. acknowledges support from the University of Liege under Special Funds for Research, IPD-STEMA Programme. M.G. acknowledges support  by the Italian Ministry for University and Research through the Levi-Montalcini program and by the Swiss National Science Foundation (SNSF) through the Ambizione program (grant PZ00P2\_174056).

\bibliographystyle{myapsrev}
\bibliography{bib}

\newpage
\appendix

\section{Comments on doping-dependent transport performance}
\label{sec:doping}

The mobility is a typical figure of merit for ``transport performance'': It depends on doping (i.e.\ the carrier density induced by field effects), and the relevant doping range might vary with the application.
Most first-principles computations of mobility are done in the zero carrier density limit $\mu_0 = \lim_{n \to 0}\mu$.
Here, instead, we focus on 2D semiconductors with high conductivity $\sigma$ at a fixed carrier density of $n/p = 10^{13}$ cm$^{-2}$. Of course, by optimizing $\sigma$ we also maximize the mobility $\mu = \frac{\sigma}{e n/p}$ at this particular density, denoted with $\mu_{13}$.
It should be highlighted that, in general, $\mu_0 \neq \mu_{13}$, and the variation of the mobility in between these two doping regimes is not obvious to predict.

The low (but finite) doping regime is very challenging to simulate realistically.
The chemical potential is below the band edge and depends strongly on temperature. One would ideally run one simulation per temperature, with an electronic smearing corresponding to this temperature and an accordingly dense grid of k-points.
Unfortunately, even room temperature corresponds to a very low electronic smearing compared to standard DFT and DFPT calculations, resulting in very dense grids of k-points and  prohibitively expensive calculations (especially when studying many materials as in this work).
This issue is usually circumvented by simulating the neutral system and computing only $\mu_0$.
Sometimes, the most obvious consequence of doping, i.e.\ screening form free carriers, is added as an analytical post-processing correction\cite{Ma2014a,Verdi2017}, but which scattering sources should be screened by free carriers, and how, has been debated \cite{Bogulsawski1977,Boguslawski1977,Boguslawski1980}.
Moreover, field effects and screening can have non-trivial consequences, as demonstrated in TMDs\cite{Sohier2019} or in graphene \cite{Sohier2017}.
In this work we choose to fully account for doping in the calculations, but at relatively high carrier density.
This allows us to perform more realistic calculations, easier to converge, with a chemical potential within the band and a well-defined Fermi surface.
We use a smearing that is large compared to room temperature in order to have accurate results with affordable k-point grids, but the ``cold'' nature of the smearing\cite{Marzari1999} allows to gets closer to room temperature conditions. The effects of smearing can be significant at low doping, when the chemical potential is in the gap; at the high doping levels considered here, however, we expect the calculations to be representative of room-temperature conditions. Note that an alternative consistent approach to smear a finite temperature Fermi-Dirac distribution has been put forward\cite{Verstraete2001}, but it is not implemented in our computational framework.

To predict the behavior of the mobility as doping decreases, one needs to account for the variation of electron-phonon interactions.
If the strength of EPIs are constant, simple models using quadratic bands and elastic scattering show that the mobility is roughly constant\cite{Ma2014a} up to fairly large dopings and then decreases. Considering inelastic scattering increases mobility at small doping  because there are no more available states for phonon emission close to the band edge. In any case, maximizing $\mu_{13}$ implies maximizing $\mu_0$.
However, for the materials considered here, EPIs are likely to be at least partially doping-dependent via free-carrier screening. Indeed, those are all single-valley materials, implying that scattering is dominated by momenta smaller than the size of the Fermi surface, where free carriers screening is efficient.
If all EPI were sensitive to free-carrier screening, we would see an opposite trend, with mobility increasing as a function of doping\cite{Ma2014a}, as free carriers screen the scattering sources. $\mu_0$ might be then significantly lower than the $\mu_{13}$ computed here. In practice, the magnitude of this trend will depend on which EPIs are sensitive to screening and how strong the bare EPIs are.

The electron-doped materials studied here (InSe, GaSe, Bi$_2$SeTe$_2$, Sb$_2$SeTe$_2$) have strong Born effective charges and Fr\"ohlich couplings, which is screening-sensitive and sharply increases at small momenta. For InSe, we compute $\mu_{13} \approx 490$ cm$^2/$Vs, compared to $\mu_{0} \approx 100$ cm$^2/$Vs computed in Refs.~\onlinecite{Li2019a,Ma2020} including polar effects and $\mu_{0} = 488$~cm$^2/$Vs when the Fr\"ohlich coupling is suppressed\cite{Ma2020} (as due to screening). Similar trends are expected for the other electron-doped materials in this work. Whether $\mu_0$ or $\mu_{13}$ is more relevant to a certain operating doping range depends on the critical carrier density at which free-carrier screening becomes efficient. In 2D and assuming a constant density of states per area $D$, one can derive $n = k_{B}T D \ln{\left[1+e^{\frac{\mu_F-\varepsilon_C}{k_{B}T}}\right]}$
where $k_{B}T$ is the thermal energy, $\mu_F$ the chemical potential (entering the Fermi-Dirac distribution)
and $\varepsilon_C$ the bottom of the conduction band. If we estimate the onset of free-carrier screening as $\mu_F-\varepsilon_C>-k_{B}T$ (when the occupations are not dominated by the tail of the Fermi-Dirac distribution), we obtain $n> 5 \times 10^{11}$ cm$^{-2}$ for all the electron-doped materials studied here.

The hole-doped materials studied here have weaker Born effective charges, but that is not to say that the remaining EPIs are not sensitive to screening; nevertheless, smaller variations of the mobility are to be expected.

\section{Additional electron-phonon scattering data}
\label{app:add_data}

\begin{figure*}[h]
 \includegraphics[width=0.9\textwidth]{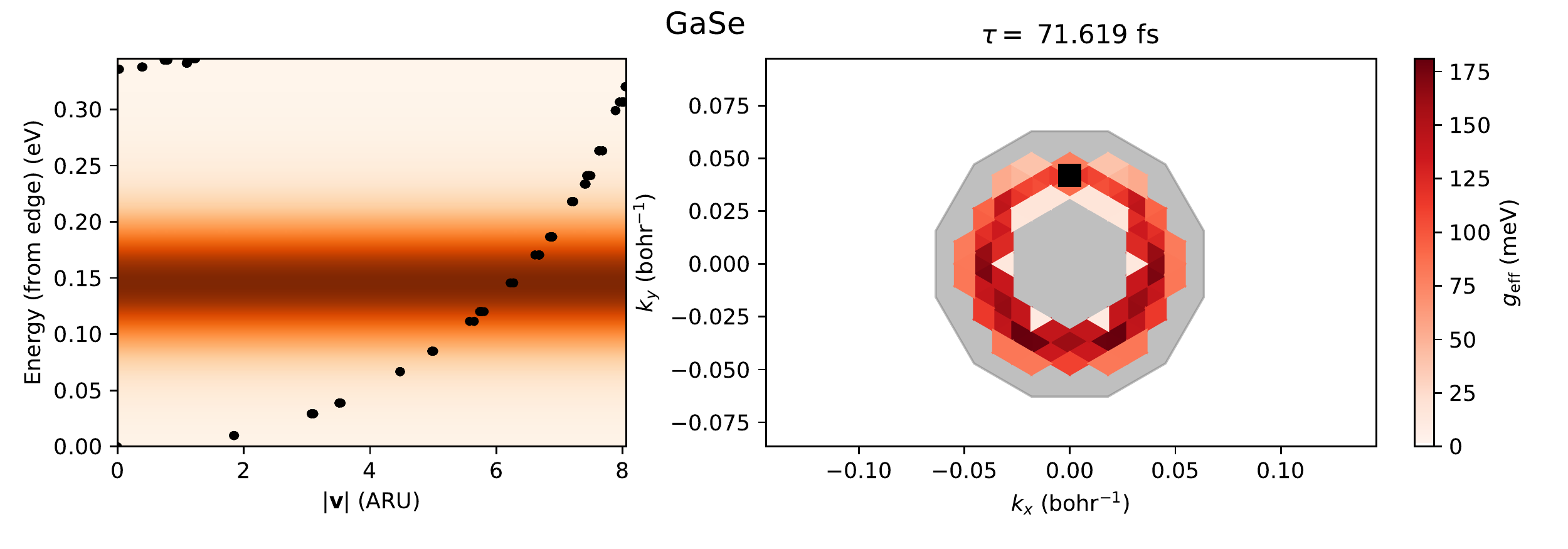}
 \includegraphics[width=0.9\textwidth]{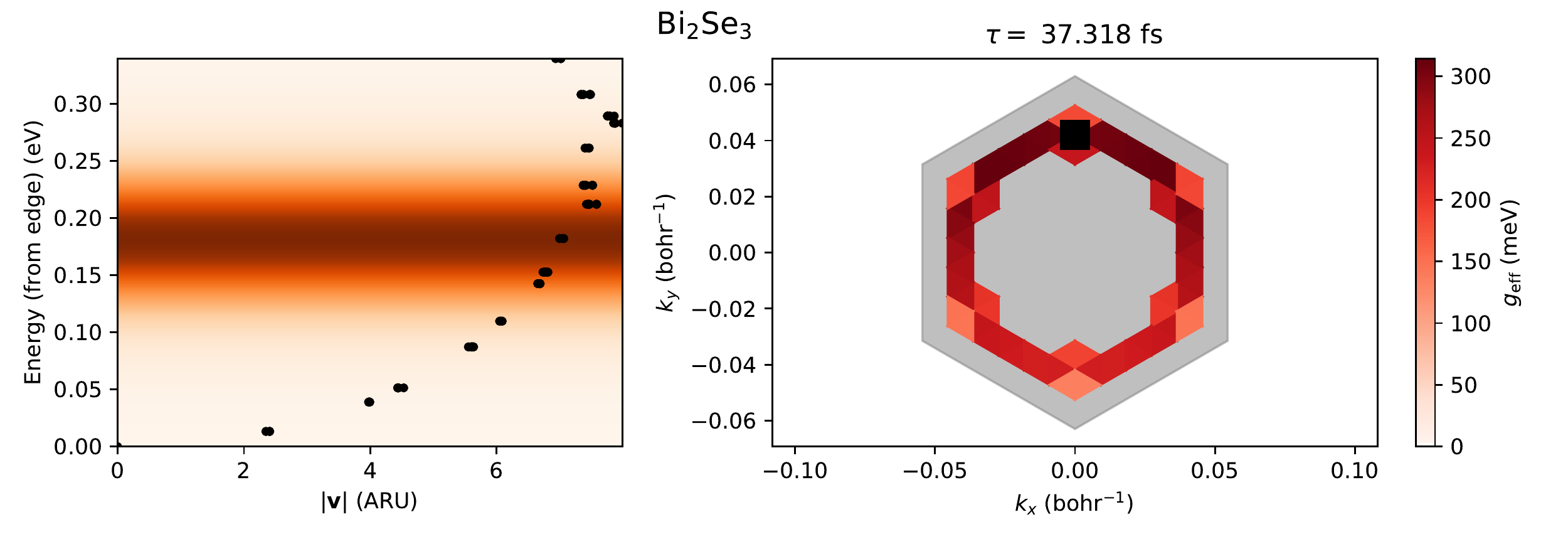}
  \includegraphics[width=0.9\textwidth]{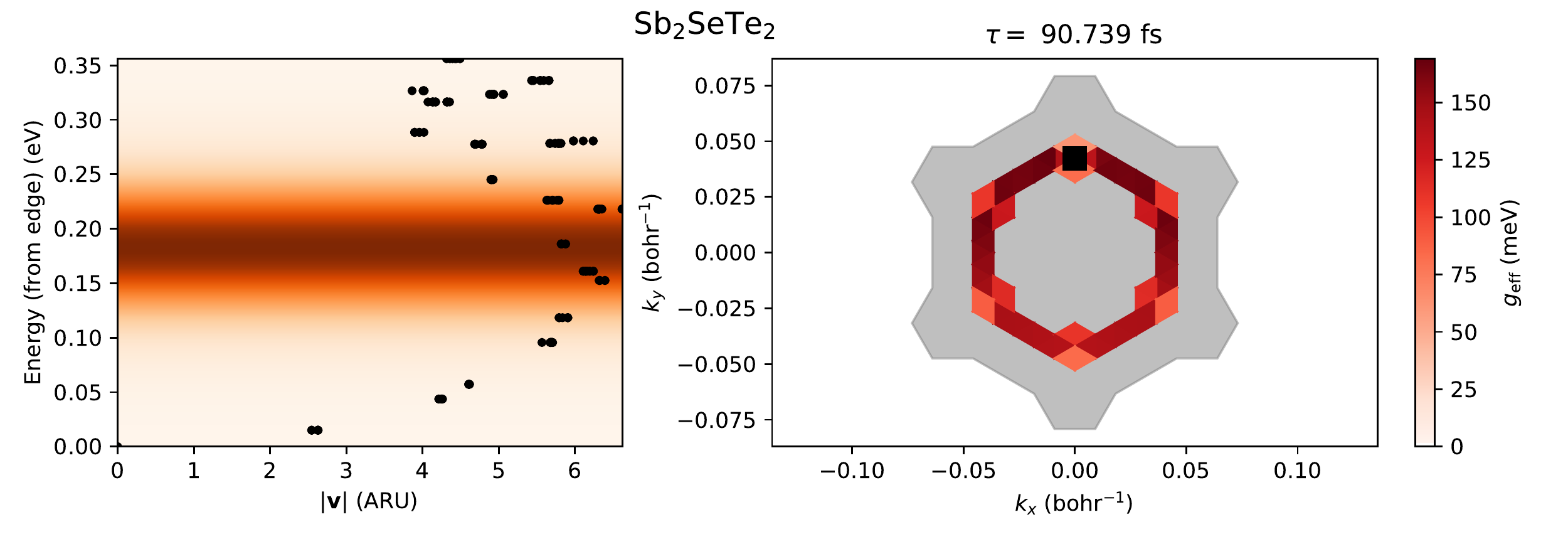}
 \includegraphics[width=0.9\textwidth]{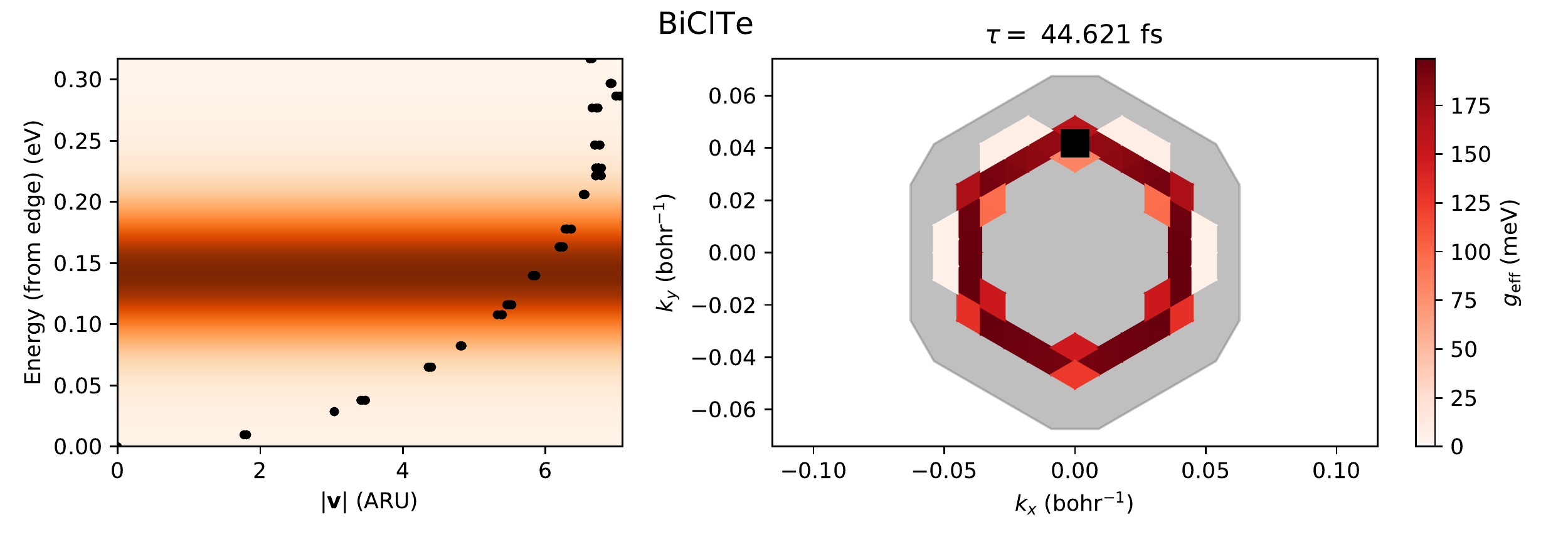}
   \caption{Velocity and scattering plots of GaSe-e,
   Bi$_2$Se$_3$-e, Sb$_2$SeTe$_2$-e, and BiClTe-e, in order of decreasing conductivity.}
   \label{fig:add_scatt_plots}
 \end{figure*}

Fig.~\ref{fig:add_scatt_plots} shows the transport properties of GaSe-e,
Bi$_2$Se$_3$-e, Sb$_2$SeTe$_2$-e, and BiClTe-e.

\begin{figure*}[h]
 \includegraphics[width=0.45\textwidth]{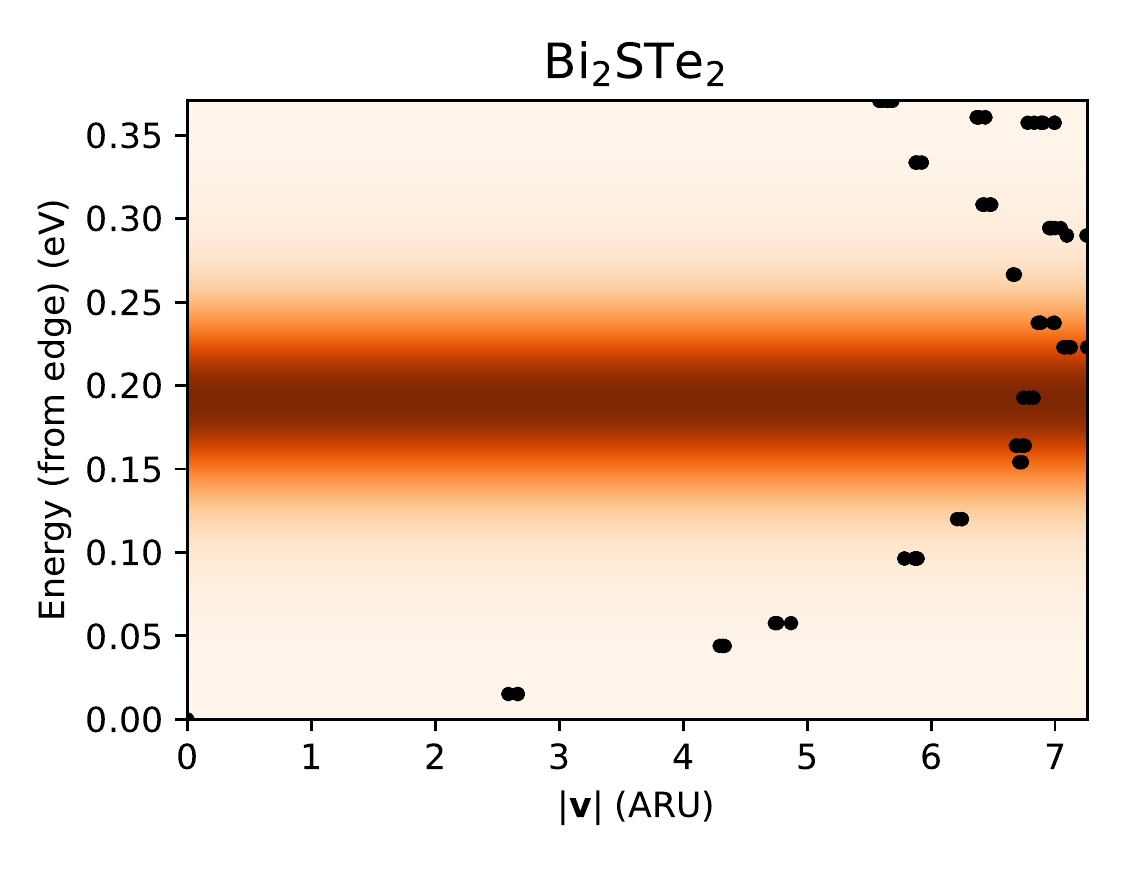}
   \caption{Velocity plot of neutral Bi$_2$STe$_2$-e. Electron-phonon scattering was not computed but it is expected to be similar to Bi$_2$SeTe$_2$-e.}
   \label{fig:velplot_BiSTe2}
 \end{figure*}

Fig.~\ref{fig:velplot_BiSTe2} shows the velocity plot for Bi$_2$STe$_2$-e, for which phonon-limited transport was not computed, but is expected to be similar to Bi$_2$SeTe$_2$-e.

\end{document}